# InP nanocrystals on silicon for optoelectronic applications

Slawomir Prucnal[1,2], Shengqiang Zhou[1], Xin Ou[1], Helfried Reuther[1], Maciej Oskar Liedke[1,3], Arndt Mücklich[1], Manfred Helm[1], Jerzy Zuk[2], Marcin Turek[2], Krzysztof Pyszniak[2] and Wolfgang Skorupa[1]

[1] Institute of Ion Beam Physics and Materials Research, Helmholtz-Zentrum Dresden-Rossendorf, PO Box 510119, D-01314 Dresden, Germany
[2] Maria Curie-Sklodowska University, Pl. M. Curie-Sklodowskiej 1, 20-035 Lublin, Poland
[3] Present address: Institute of Radiation Physics, Helmholtz-Zentrum Dresden-Rossendorf, PO Box 510119, D-01314 Dresden, Germany.

**Abstract**

One of the solutions enabling performance progress, which can overcome the downsizing limit in silicon technology, is the integration of different functional optoelectronic devices within a single chip. Silicon with its indirect band gap has poor optical properties, which is its main drawback. Therefore, a different material has to be used for the on-chip optical interconnections, e.g. a direct band gap III–V compound semiconductor material. In the paper we present the synthesis of single crystalline InP nanodots (NDs) on silicon using combined ion implantation and millisecond flash lamp annealing techniques. The optical and microstructural investigations reveal the growth of high-quality (100)-oriented InP nanocrystals. The current–voltage measurements confirm the formation of an n–p heterojunction between the InP NDs and silicon. The main advantage of our method is its integration with large-scale silicon technology, which allows applying it for Si-based optoelectronic devices.

## 1. Introduction

Nowadays, CMOS at the 45 nm node being in production for a couple of years has been replaced by 32 nm, and soon CMOS at the 22 nm node will be in production. The further downsizing of CMOS devices below 16 nm depends on overcoming of the practical limits caused by the integration issues, such as chip performance, cost of development and production, power dissipation, reliability, etc. The integration of III–V compound semiconductors with silicon is unavoidable for a further performance enhancement of optoelectronic devices. Light emitting compound semiconductor quantum structures giving the possibility to tune the luminescence emission from the ultraviolet (UV) up to the infrared (IR) are much more attractive than those made of silicon [1]. The III–V compound semiconductors have a huge potential as non-silicon transistor channel materials for future high-speed and low-power logic CMOS applications [2–5]. The open question is which technology integrating III–V compound semiconductors with silicon will be used.

Conventionally, the integration of III–V semiconductors with silicon is based on heteroepitaxial growth of multi-layered structures on silicon or direct wafer bonding technology [6]. Different compound semiconductors (quantum dots, nanowires, thin films) heterogeneously integrated onto a silicon substrate have been intensively studied [7, 8]. Devices made of such structures combine the high carrier mobility and high luminescence efficiency of III–V semiconductors with the well-developed silicon technology [9]. Recently, by using a low-temperature direct wafer bonding process, high-performance lasers, amplifiers, photodetectors and modulators have been demonstrated on a hybrid silicon platform [10]. Another approach towards the integration of binary semiconductors with silicon focuses on epitaxial lift-off methods. The epitaxial lift-off method was used to transfer an ultrathin InAs layer onto a Si/SiO$_2$ substrate in order to fabricate FETs [11]. This technique is very promising and can probably be used for different types of semiconductors, but it is still limited to small areas and suitable only for SOI technology.

Up to now the modification of the electronic and optoelectronic properties of silicon in the microelectronic industry is based on the ion implantation method and subsequent thermal annealing. An alternative technique to the epitaxial growth of III–V nanostructures is high fluence ion implantation. Potentially, any kind of compound semiconductor can be formed in any solid substrate with this method. The post-implantation thermal annealing restores the initial properties of the matrix and leads to the formation of nanocrystals from supersaturated solid solutions. Recently we have synthesized inverted crystalline InAs nanopyramids (NPs) at the SiO$_2$/Si interface [12] and on Si fingers by ion implantation, flash lamp annealing and wet chemical etching [13]. Measurements demonstrate that the n-type InAs NPs are degenerate with a type-II band alignment of the InAs–Si. In this paper we propose a compact, CMOS compatible solution for the integration of indium phosphide compound semiconductors with silicon technology for optoelectronic applications. The optoelectronic properties of InP are very attractive for wide applications in optoelectronics. Besides near-infrared band gap emission at around 920 nm, InP crystals can be used as THz emitters and detectors. The experimental results obtained by means of cross-section TEM, scanning electron microscopy (SEM), x-ray diffraction (XRD) and optical methods (μ-Raman and photoluminescence spectroscopy) proves the formation of high-quality InP nanocrystals. Moreover, the current–voltage (*I*–*V*) characteristics confirm the p–n heterojunction formation between the n-InP and p-type silicon substrate.

## 2. Experimental

Single crystalline p-type (100) silicon wafers with a resistivity in the range of 1–20 Ω cm covered by a 60 nm thick thermally grown oxide layer were implanted sequentially with P and In ions at a depth of ~80 nm from the sample surface. Initially, the 70 keV P$^+$ ions were implanted with a dose of $3 \times 10^{16}$ ion cm$^{-2}$. Subsequently the samples were implanted by 170 keV In$^+$ ions with a dose of $2 \times 10^{16}$ ion cm$^{-2}$. Both implantations were performed at room temperature. All the high-dose implantations were done employing an efficient ion source with an internal evaporator [14]. The expected In and P ion concentration is in the range of $5 \times 10^{21}$ cm$^{-3}$ in p-Si. The distribution of the implanted ions was calculated by the SRIM 2007 code [15]. After the implantation the samples were annealed for 20 ms by flash lamp annealing (FLA) with 600 or 700 °C preheating for 1–5 min and the annealing temperature during FLA ranging from 900 up to 1350 °C. In all cases the samples were annealed in argon ambient. For the InP/Si heterojunction formation, wet chemical etching with 30% KOH solution or reactive ion etching (RIE) using a mixture of SF$_6$ and O$_2$ plasma was applied. Independently on the final etching type, in a first step the SiO$_2$ layer was removed with a HF:H$_2$O solution. The selective wet chemical etching was performed at room temperature with an etching rate of about 25 nm min$^{-1}$. For the RIE etching a 200 W plasma with SF$_6$ and O$_2$ gas flows of 64 and 17 sccm was used, respectively. After 1 min of the RIE etching approximately 700 nm of silicon was selectively removed. The microstructural properties of InP nanocrystals were investigated by high-resolution transmission electron microscopy (HRTEM), high-angle annular dark-field scanning TEM (HAADF-STEM), energy-dispersive x-ray spectroscopy (EDS) in a cross-sectional geometry by means of an FEI Titan 80-300 STEM operating at 300 keV and x-ray diffraction (XRD). XRD was performed using a Siemens D5005 diffractometer with a Cu-target source. The setup was equipped with a Göbel mirror to enhance the brilliance. The elemental distribution on the etched samples was measured by μ-Auger spectroscopy with lateral resolution of about 15 nm. The optical properties were investigated by μ-Raman and photoluminescence spectroscopy. The μ-Raman spectra were recorded at room temperature in a backscattering geometry in the range of 150–600 cm$^{-1}$ using a 532 nm Nd:YAG laser. The same type of laser was used for the PL excitation. For the *I–V* measurements conductive atomic force microscopy (C-AFM) equipped with the PtIr coated n$^+$-Si cantilever tip was used. The surface morphology after each preparation step was characterized by means of the atomic force microscopy (AFM).

## 3. Results and discussion

### 3.1. Microstructural properties of InP nanocrystals

The evolution of the InP nanocrystal growth during flash lamp annealing and the influence of the annealing parameters on the crystallographic orientation, shape and size of InP nanocrystals were investigated using bright-field HRTEM by keeping the electron beam along the [011]-zone axis. Figure 1 shows cross-section TEM images of FLA prepared samples with different preheating parameters. The flash lamp annealing performed on implanted samples at a temperature lower than the melting point of bulk InP (1062 °C) did not lead to crystalline InP nano-inclusions. After annealing at 1100 °C for 20 ms with preheating at 600 °C for 1 min (sample A) InP nanoparticles with an average size of 60 nm, triangular in the sectional view, were observed (see figure 1(a)). The InP nanoparticles are polycrystalline with amorphous InP inclusions. Moreover, some of them contain inclusions of crystalline indium which did not react with phosphorous. The increase of the FLA temperature up to 1300 °C with preheating at 600 °C for 1 min (sample B) leads to the formation of zinc-blende hemispherical single crystalline (111)-oriented InP nanocrystals (see figure 1(b)). However, some of them are still decorated with indium nanocrystals. The electron diffraction pattern clearly indicates (111)-oriented InP nanocrystals. The interplanar distances measured in the selected area are 0.170 and 0.341 nm, which is 0.6% and 0.9% larger than the interplanar distance for bulk InP with $d_{222} = 0.169$ and $d_{111} = 0.338$ nm, respectively [16]. The slightly higher value (compared to bulk InP) of the interplanar distances suggests a small tetragonal distortion of the zinc-blende InP nanocrystals. Figure 1(c) shows the TEM image of InP nanocrystals obtained after FLA at 1300 °C with 600 °C preheating for 3 min (sample C). An increase of the preheating time up to 3 min significantly improves the nanostructural properties of the InP NCs. First of all, the InP NCs are fully relaxed and are (100)-oriented. Within the TEM investigated area we did not find any metallic indium, which suggests that all the indium is incorporated into nanocrystals. By increasing the preheating time the average diameter of InP NCs increases up to 300 nm. A Moiré pattern is visible on the periphery of the hemispherical InP NC. Moiré fringes are formed by superposition of two crystalline layers with different lattice parameters: the InP nanocrystal on the top of Si with lattice parameters of 0.5868 nm and 0.5431 nm, respectively.

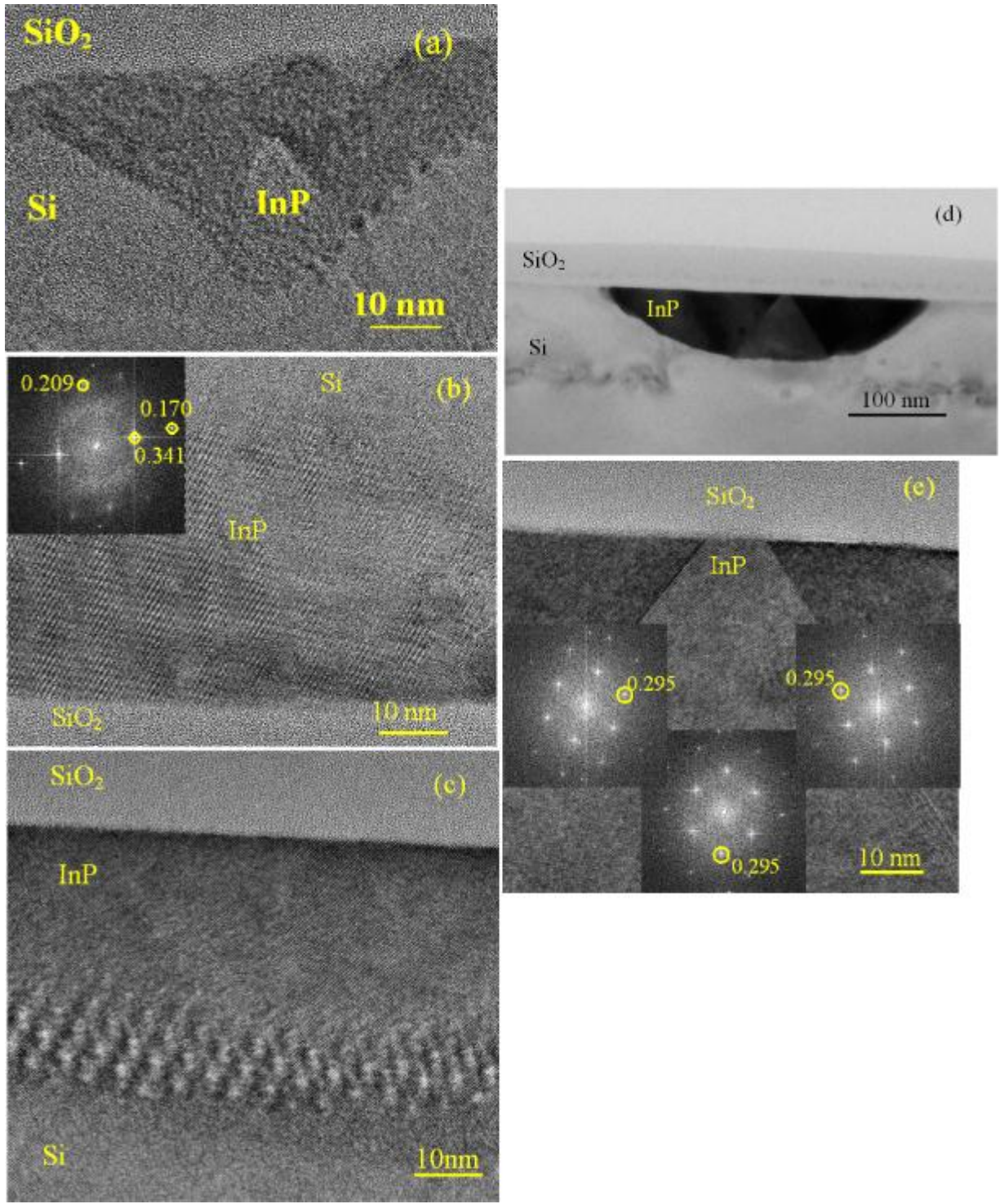

**Figure 1.** High-resolution cross-section TEM micrographs of Si co-implanted by P and In followed by flash lamp annealing for 20 ms with different preheating times and temperatures: (a) sample A: preheating −600 °C for 1 min with FLA at 1100 °C, (b) sample B: 600 °C for 1 min with FLA at 1300 °C, (c) sample C: 600 °C for 3 min with FLA at 1300 °C, (d) sample D: 700 °C for 2 min with FLA at 1200 °C and (e) bright-field HRTEM micrograph of the InP nanocrystal from (d). The insets in (b) and (e) show selected area electron diffraction patterns.

Raising the preheating temperature up to 700 °C followed by the FLA at 1200 °C (sample D) enlarges the average size of the InP NC to the range from 400 up to 500 nm (see figure 1(d)). The InP nanoparticles consist of (100)-oriented crystals located in the middle and (111)-oriented ones in the surrounding. The interplanar distances measured for the (111)-oriented part of the nanocrystals are 0.172 and 0.343 nm, while the (100)-oriented part shows 0.295 and 0.338 nm, the same as for sample C. The obtained values of interplanar distances for (111)-oriented nanocrystals are slightly higher than for sample B, which suggests slightly higher distortion. In case of the (100)-oriented part of nanocrystals the obtained value of interplanar distances matches exactly the value for bulk (100) InP [17]. Based on the microstructural investigation of the implanted and annealed samples we have shown that it is possible to control the crystallographic orientation and the size of the InP nanocrystals just by varying the annealing parameters. Generally, high-quality InP NCs are obtained after millisecond annealing at temperatures at least 100 °C higher than the melting point of bulk InP. During the preheating the silicon amorphized by ion implantation recrystallizes, and indium and phosphorus form agglomerates, which are recrystallized later on during the flash. Hence, the preheating parameters influence mainly the size of the InP nanoparticle while the FLA parameters control the crystallographic orientation of the InP NCs.

In a first attempt, the composition of the InP NCs was investigated by energy-dispersive x-ray spectroscopy in the cross-sectional geometry. Figure 2 shows the EDS spectra obtained from the samples A–D measured in the middle of the nanocrystals presented in figures 1(a)–(d). We can clearly see that all of the nanocrystals are composed of indium and phosphorous atoms with silicon inclusions.

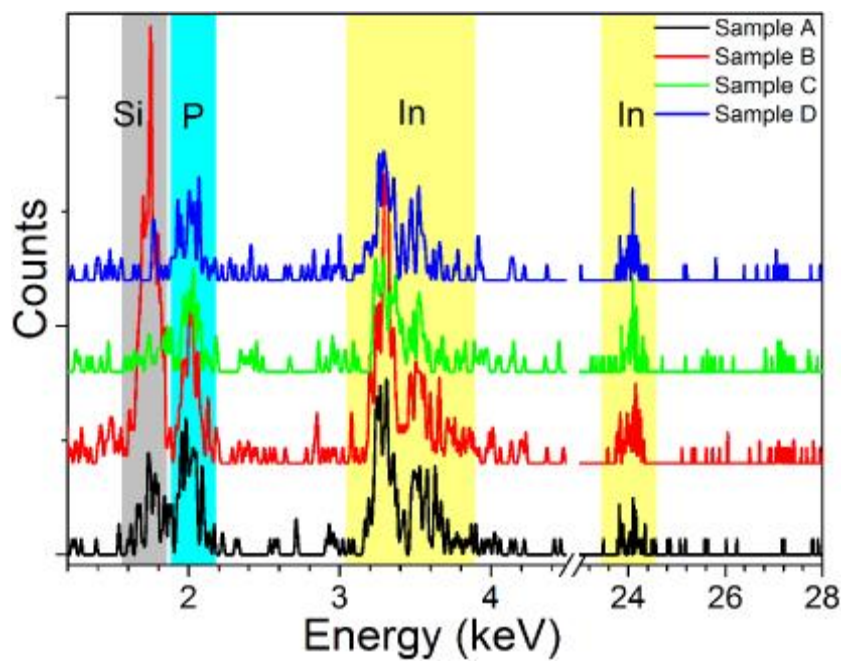

**Figure 2.** EDS spectra of the elemental composition measured in the middle of each InP nanocrystal presented in figure 1. The spectra are vertically shifted for clarity.

The chemical composition of InP NCs was calculated based on the EDS results. Taking into account the different sensitivity of the system for different elements we have found that the chemical composition of the InP NCs did not change significantly from sample to sample and the InP NCs are composed of $52 \pm 2\%$ of In, $47 \pm 2\%$ of P and about 1% of silicon. The exception was sample B, where the silicon-related peak dominates the spectrum. Since we are dealing with InP nanocrystals embedded into the silicon matrix the strong signal from silicon may come from the substrate. The silicon contamination can be taken as the advantage since silicon acts as an n-type dopant for most III–V compound semiconductors. Having in mind the application aspect of the obtained nanostructures we have developed a method for selective chemical etching based on the reagents commonly used in the microelectronic industry [13]. The main aim of this part of research was to fabricate nanostructures which would allow us to perform current–voltage ($I$–$V$) measurements. Such measurements could directly confirm the heterojunction formation between the n-type InP nanocrystals and the p-type silicon substrate. Based on the high-resolution cross-section TEM results, the best features are revealed in sample C, annealed at 1300 °C for 20 ms with preheating at 600 °C for 3 min. Therefore, for the next experiments, samples annealed at such conditions were considered. After removing the $SiO_2$ capping layer in HF:$H_2O$ solution the silicon was selectively etched in 30% KOH at room temperature. The etching rate of silicon at such conditions was estimated to be about 25 nm min$^{-1}$, which is about 5 nm min$^{-1}$ faster than for the InAs–Si system [13]. The topography and superficial distribution of silicon, indium and phosphorous after the annealing and the selective chemical etching were investigated by AFM and μ-Auger spectroscopy, respectively.

Figure 3(a) shows the AFM topography of the sample annealed and selectively etched for 6 min. The size of InP nanocrystals varies from 300 up to 450 nm with a height of 150 nm. During the selective etching the smallest nanocrystals are removed from the surface due to under-etching. Hence, if the etching is prolonged up to 10 min (the InP–Si columns are 250 nm high) only the biggest nanocrystals remain on the surface, which significantly improves the size distribution of the InP–Si heterostructures. Figures 3(b)–(d) show the superficial distribution of the silicon, phosphorous and indium, respectively, obtained from the flash lamp annealed sample etched for 6 min. The colours indicate different concentrations of elements at the surface: bright yellow corresponds to the high concentration. As can be seen in figure 3(b), the area covered by InP nanocrystals reveals low silicon content while the same spots in figures 3(c) and (d) show high P and In concentrations. Hence, we may conclude that the structures visible on figure 3 are composed of indium and phosphorous. The position of bright spots in figure 3(d) is slightly shifted compared to figures 3(b) and (c) due to the small sample drift during the measurements.

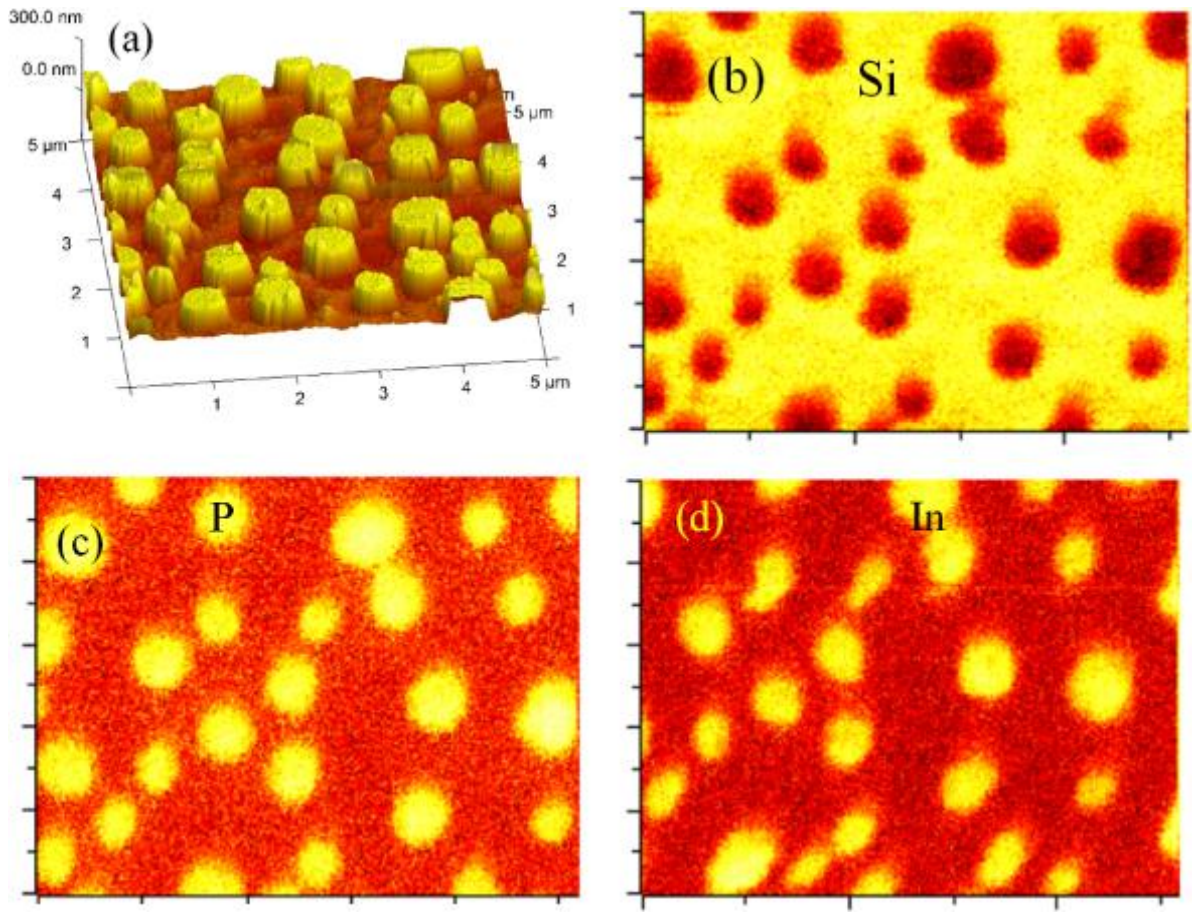

**Figure 3.** AFM topography (a) and superficial distribution of Si (b), P (c) and In (d) obtained by μ-Auger spectroscopy from the annealed and 6 min etched sample. Yellow areas correspond to a high concentration of the investigated element.

The crystallographic orientation of the sample C after 10 min chemical etching was determined by x-ray diffraction (see figure 4). The x-ray spectrum reveals two main 2θ peaks at 32.95° and 69.17°, corresponding to the x-ray diffraction on a (100)-oriented silicon substrate.

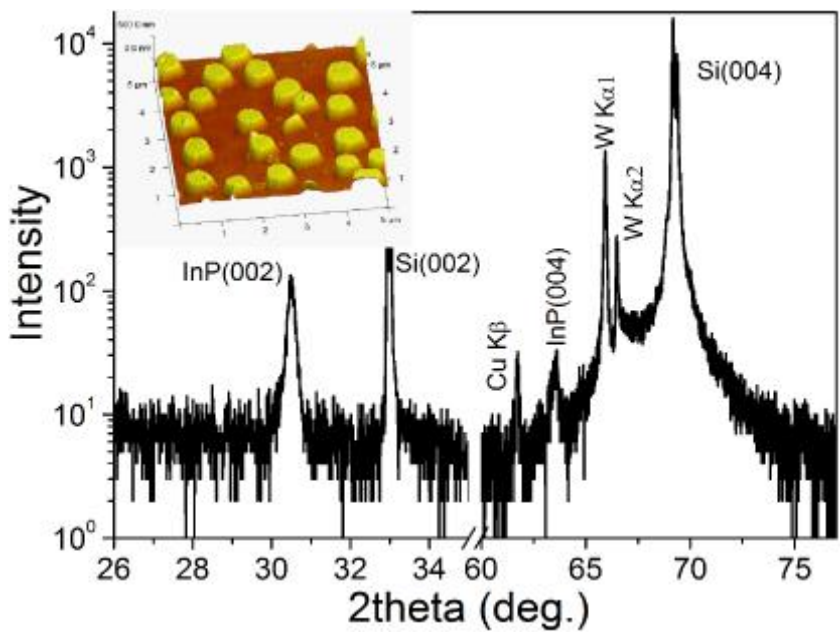

**Figure 4.** Diffraction pattern of InP–Si heterostructures obtained from the sample annealed at 1300 °C for 20 ms with preheating at 600 °C for 3 min. The sample was selectively etched in KOH solution for 10 min. The inset shows the AFM topography.

Besides the silicon-related peaks the x-ray diffraction spectrum reveals two peaks at 30.47° and 63.57° (2θ) which are typically observed from zinc-blende (100)-oriented monocrystalline InP [18, 19]. The full width at half maximum for the InP$_{(002)}$ peak is 0.2° (2θ). The x-ray diffraction pattern results are consistent with the cross-section TEM imaging of the sample C, where only (100) InP nanocrystals were found. The inset in figure 4 shows the AFM topography obtained from the same sample after 10 min etching in KOH solution. The height of the InP–Si heterostructures is about 250 nm, which exactly matches the estimated etching rate after 10 min. The average diameter of the InP nanocrystals is about 420 ± 30 nm. The InP nanocrystals with diameters below 300 nm are removed from the surface during the etching process.

### 3.2. Optoelectronic properties of the InP–Si heterostructures

Since we are dealing with InP nanocrystals on a silicon wafer, the Raman spectra contain silicon-related phonon modes. The main silicon phonon mode is located at 520.5 cm$^{-1}$, which we used as the reference for the peak position calibration in all spectra. Besides the TO phonon mode, the Raman spectra obtained from the crystalline silicon reveal a peak at 302.7 cm$^{-1}$ due to the second-order two-transverse acoustic phonon (2TA$_{Si}$) scattering (see solid line in figure 5) [20].

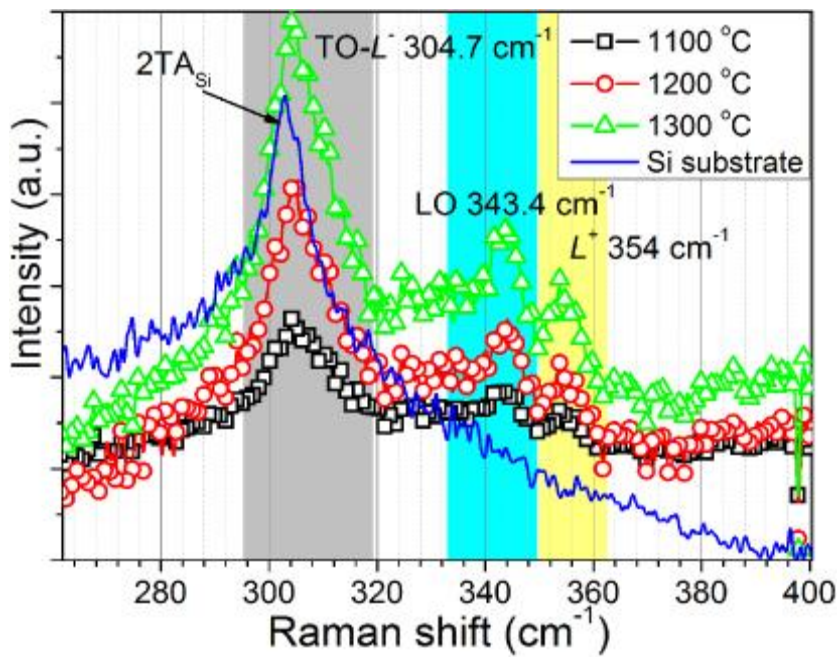

**Figure 5.** μ-Raman spectra of flash lamp annealed sample at different temperatures for 20 ms with preheating at 600 °C for 3 min after 4 min selective etching. The Raman spectrum from a Si substrate (non-implanted, but FLA at 1300 °C) is shown for comparison as well. The spectra are vertically shifted for clarity.

The position of 2TA$_{Si}$ overlaps with the TO phonon mode from InP. In order to minimize the influence of the substrate signal on the data interpretation, a non-implanted silicon wafer was annealed at the same conditions as the implanted one. Next, the Raman spectra related to the non-implanted and annealed silicon wafers were subtracted from the spectra obtained from the implanted samples that were thermally treated in the same way, but contained InP nanocrystals. Figure 5 shows the µ-Raman spectra of the implanted, flash lamp annealed and 4 min etched samples. The Raman spectrum of non-implanted flash lamp annealed sample at 1300 °C composed of 2TA$_{Si}$ phonon mode is shown as well. The Raman spectra consist of three main peaks located at about 305, 343 and 354 cm$^{-1}$, corresponding to transverse optical (TO) combined with LO phonon–plasmon coupled modes (LOPCM)—L$^-$ coupled mode, longitudinal optical (LO) phonons and L$^+$ coupled mode observed from crystalline InP, respectively [21, 22]. According to our microstructural investigation, samples preheated at 600 °C for 3 min and subsequently FLA treated at temperatures up to 1200 °C contain (100)- and (111)-oriented InP nanocrystals, while the FLA at 1300 °C reveals single crystalline (100)-oriented InP nanoparticles only. Taking into account the selection rules for the Raman backscattering geometry from (100)-oriented zinc-blende InP, the TO phonon mode is forbidden. Hence, after the high-temperature annealing in such a sample the LO mode only should be visible. But usually even high-quality (100) InP crystals show weak TO peaks in the spectra due to some imperfections, contamination or deviations from the ideal backscattering geometry.

Moreover, the peak position of the phonon mode related to the L$^-$ branch of the LOPCM moves down in frequency towards the TO peak in heavily n-type doped InP. For an electron concentration in the range of $1 \times 10^{18}$ cm$^{-3}$ or higher, the L$^-$ mode overlaps the TO phonon mode [21]. Each sample shows also a peak at around 354 cm$^{-1}$. In the literature the phonon mode located at around 354 cm$^{-1}$ is assigned either to the surface phonon mode at the Γ point in (110) InP or to the L$^+$ branch of LOPCM in n-InP [22, 23]. Neither XRD nor cross-section TEM reveals formation of the (110) InP, but EDS measurements reveal silicon contamination in the InP nanocrystals. If we assume that the silicon incorporated into InP nanocrystals is electrically activated during liquid phase epitaxy growth of InP NCs (the annealing was performed at a temperature higher than the melting point of bulk InP), Raman peaks located at 304.7 and 354 cm$^{-1}$ in the sample C annealed at 1300 °C correspond to L$^-$ and L$^+$ modes, respectively. In the case of samples annealed at lower temperatures the 304.7 cm$^{-1}$ peak is composed of TO and L$^-$ phonon modes, since these samples exhibit (111)- and (100)-oriented InP. Hence, we can conclude that the InP nanocrystals are n-type with electron concentrations in the range of $10^{18}$ cm$^{-3}$.

Figure 6 shows the room temperature PL spectra obtained from the sample C annealed at 1300 °C for 20 ms with preheating at 600 °C for 3 min after selective chemical and reactive ion etching. For the RIE the SF$_6$/O$_2$ gas mixture was used. After chemical etching the height of the heterostructure is in the range of 250 nm, while the RIE produces about 800 nm high nanocolumns. Both etching methods are neutral to the InP nanocrystals. Independently of the etching type the PL spectra are composed of a main broad band with the maximum intensity at about 1130 nm and the peak at 920 nm. The main PL peak corresponds to the band-to-band transition in silicon, while a 920 nm peak could be assigned to band gap emission from InP nanocrystals. The power dependence of the PL reveals the typical blue shift of the InP-related luminescence, which directly confirms the origin of this emission. The blue shift of the InP band gap emission is about 30 meV per decade of the excitation laser power [24]. The chemically etched samples show more pronounced PL emission than the RIE etched, due to less surface damage. In fact the band gap luminescence obtained from InP nanocrystals is much weaker than that from silicon. It should not be surprising, since we are working with heavily n-type doped InP nanocrystals. According to the μ-Raman data the electron concentration in InP NDs is in the range of ×10$^{18}$ cm$^{-3}$. A high doping level is known to degrade the optical properties of different semiconductors due to strong non-radiative Auger recombination. We cannot exclude the existence some point defects within the nanocrystals, which may acts as non-radiative photoluminescence decay centres as well. The insets of figure 6 show the SEM micrographs obtained from KOH (a) and RIE (b) etched samples. As can be seen, the KOH etching leaves a smooth silicon surface between InP nanoparticles, while a type of black silicon was formed during the RIE. Since energetic ions bombard the surface of InP during the RIE, nanocrystals and some point defects at the surface can be generated. This phenomenon enhances the surface recombination process and reduces the PL efficiency [25]. So far the band gap alignment between InP nanocrystals or nanowires and silicon substrate is not clear [26]. The power-dependent PL is quite often used to study the band alignment of the semiconductor heterostructure. On one hand, the blue shift of the PL emission at around 920 nm confirms our assumption that the InP NDs are responsible for it. On the other hand, the relatively strong blue shift of the 920 nm emission with increasing power excitation can be evidence for the type-II band alignment between n-InP NDs and p-Si [27]. Typically, the blue shifts of the PL peak intensity with increasing excitation power density for type-II structures is due to the band bending that occurs at the interfaces between two different semiconductors.

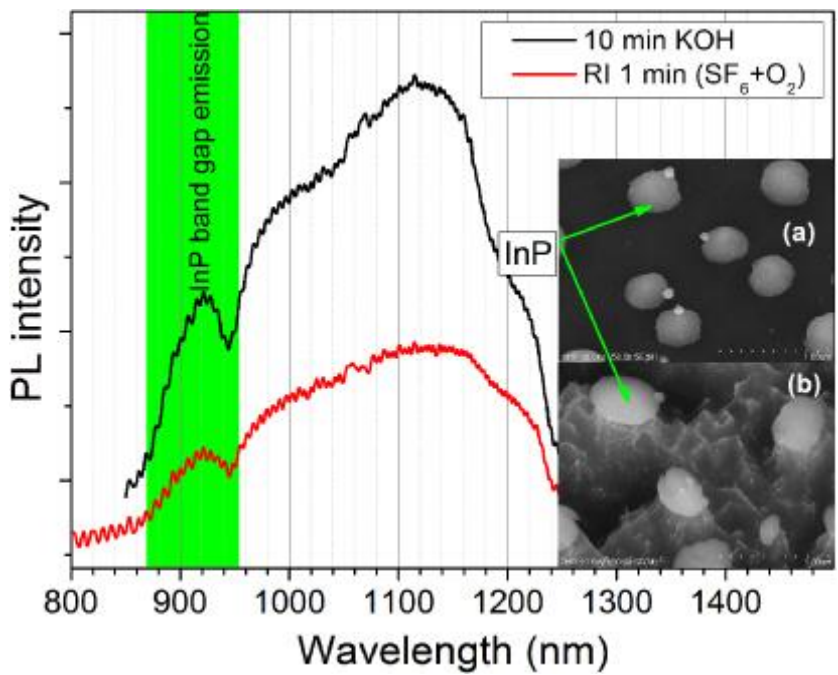

**Figure 6.** Room temperature PL spectra of annealed and selectively etched samples, either with KOH solution for 10 min or by reactive ion etching for 1 min. The insets show the SEM micrograph of a sample etched with KOH (a) and by reactive ion etching (b).

The current–voltage characteristic of the n-InP/p-Si heterojunction obtained by means of ion implantation, flash lamp annealing and selective chemical etching for 4 min is shown in figure 7. A 180 nm thick gold layer was sputtered on the back side of p-Si for the electrical investigation. The upper parts of the heterostructures were connected to the heavily doped n-type silicon cantilever coated with 20 nm PtIr during c-AFM measurements. The bias voltage was applied in the reverse ($V < 0$) and forward direction ($V > 0$) between the back gold contact and the top of the InP nanocrystal. The obtained I–V curves exhibited a typical diode behaviour. The presented p–n heterojunction shows a relatively large reverse bias leakage current due to some point defects at the silicon/InP NDs interface. The reverse bias leakage current can be reduced by decreasing the p–n junction area and improvement of the silicon/InP NDs interface quality. Decreasing of the p–n junction area can be obtained either by prolonged selective etching or using smaller InP nanocrystals. While the quality of the silicon/InP interface can be improved most probably by using long-term low-temperature annealing performed after the p–n junction formation. An influence of the silicon/InP interface properties on the quality of the n-InP/p-Si heterojunction are under investigation. The insets in figure 7 show the topography (a) and current response from the sample for electron injection in the reverse (b) and forward direction (c). Before the measurements the sample was exposed to an oxygen plasma in order to form a thin barrier oxide layer on silicon. After plasma treatment the complete sample surface was oxidized. Then, the oxide from the InP nanocrystals was chemically etched with HCl, which is neutral to the silicon dioxide. The colour contrast describes the scale: the brightness of colours corresponds to the height of the nanostructure presented in inset (a) or current level in insets (b) and (c). For the current response measurements both in the reverse and forward direction a voltage of +2 and −2 was applied to the tip, respectively. As can be seen, in the reverse direction no current flow was detected, independently of the measurement area, while in the forward direction the current was observed only when the tip crosses the InP nanocrystals. It directly proves the n–p heterojunction formation between n-type InP nanocrystals and the p-type silicon substrate.

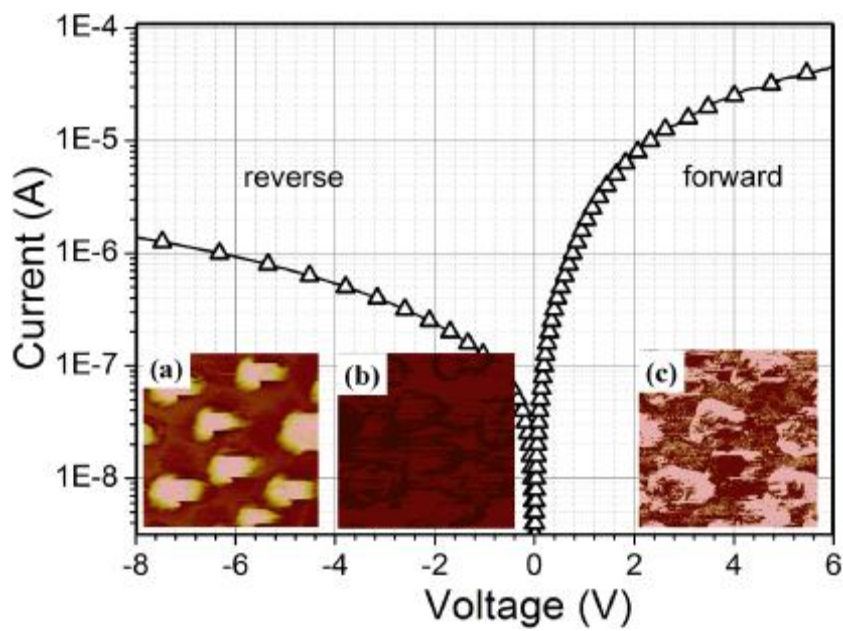

**Figure 7.** Semilogarithmic current–voltage characteristic of the n-InP/Si heterojunction. The insets show topography (a) and current response from the sample for electron injection in the reverse (b) and forward direction (c) of the n-InP/p-Si system received by C-AFM.

## 4. Conclusions

n-InP/p-Si heterostructures were successfully fabricated by sequential ion implantation, millisecond range flash lamp annealing and selective chemical etching. The monocrystalline (100) InP nanoparticles are formed during a high-temperature short-pulse annealing due to liquid phase epitaxy. Both the optoelectrical and microstructural properties of the presented system confirm the formation of high-quality n–p heterojunctions, which can be applied for the silicon based optoelectronic devices and nanosized power supplies. The presented results open a new road for the integration of the III–V semiconductors with silicon technology.

## Acknowledgments

The authors would like to thank H. Hilliges and G. Schnabel from HZDR for careful sample preparation. This work was partially supported by the Polish Ministry of Science and Higher Education, Grant No N N515 246637 and the Helmholtz-Gemeinschaft Deutscher Forschungszentren (HGF-VH-NG-713).